\documentclass[a4paper]{jpconf}
\usepackage{graphicx}
\usepackage{epstopdf}
\newcommand{\Pb}{Pb-Pb $\sqrt{s_{NN}} = 2.76$ TeV }

\begin{document}
\title{Event by event di-hadron correlations in Pb-Pb 2.76 TeV collisions from the ALICE experiment.}

\author{Anthony R. Timmins}

\address{Department of Physics, University of Houston, 617 Science and Research Building 1, Houston, TX 77204}

\ead{anthony.timmins@cern.ch}

\begin{abstract}
The large multiplicities at the LHC may permit flow harmonics to be determined on an event by event basis in Pb-Pb collisions. We extract these harmonics from event by event di-hadron correlations. Within a fine centrality bin, we find the correlation function varies substantially on an event by event basis, indicating large fluctuations in the initial conditions for a given impact parameter. Such large fluctuations lead to some events being highly triangular or highly elliptical, where the angular correlation function is completely dominated by the respective second and third Fourier harmonics. We will show unfolded $v_{2}$ distributions for various centralities, and implications for our understanding of the initial conditions.

\end{abstract}

\section{Introduction}

Relativistic heavy-ion collisions aim to create the QGP (Quark-Gluon Plasma), a unique state of matter where quarks and gluons can move freely over large volumes in comparison to the typical size of a hadron. Spatial inhomogeneities in the initial state can create pressure gradients in the QGP, which induce angular correlations known as azimuthal flow. It is convenient to express the single particle density of produced particles as a Fourier series \cite{flow1}:
\begin{equation} 
\label{eq:vnvn}
\frac{dN}{d\phi} \propto 1 + 2\sum_{n=1}^{\infty} v_{n} cos (n(\phi-\psi_{n}))
\end{equation}  
where $v_n$ and $\psi_{n}$ are the magnitude and direction of order ${n}$ harmonic flow. The magnitude $v_n$ is typically extracted from azimuthal correlations averaged over an event ensemble, where the second order (elliptic flow) is usually the dominant contribution. It has been proposed that event by event fluctuations in the initial state give rise to the odd orders \cite{fluc}. They also lead to variance in $v_{n}$ ($\sigma_{vn}$) for all orders, that can be extracted from data via two and four particle cumulants \cite{cum}. These fluctuations should be also evident in the event-wise pair distributions. All orders of $v_n$ can be extracted without the need to identify $\psi_{n}$ using the pair density:
\begin{equation} 
\label{eq:vN}
\frac{dN}{d \Delta \phi} \propto 1 + 2\sum_{n=1}^{\infty} V_{n\Delta} cos (n\Delta \phi)
\end{equation}  
where $\Delta \phi$ is the angular difference between the two particles in a pair. For factorization, it is important that the only correlation between the two particles is due to flow. In this case, $V_{n\Delta}$ factorizes into the product of $v_n$, $V_{n\Delta} = v_{n}^2$. Factorization has been demonstrated for these energies at low $p_T$ \cite{HadDe}. Measurements of di-hadron correlations from ALICE show $v_{2}$ and $v_{3}$ manifest themselves as a nearside "ridge" and awayside "double hump" for very central collisions \cite{HadDe,HigherOrder,MeUT}. 
We present measurements of event by event di-hadron correlations in \Pb collisions. The large multiplicities at the LHC and near full $p_{T}$ acceptance of the ALICE detector favour such measurements. Values of $V_{2\Delta}$ and $V_{3\Delta}$ are extracted from these distributions, and we will show events with large values of $V_{2\Delta}$ and $V_{3\Delta}$. Measurements of event wise $V_{2\Delta}$ allow for the full $v_2$ distributions to be obtained, which will also be shown. 

\begin{figure}[t]
\begin{tabular}{ll}
\includegraphics[width = 0.455\textwidth]{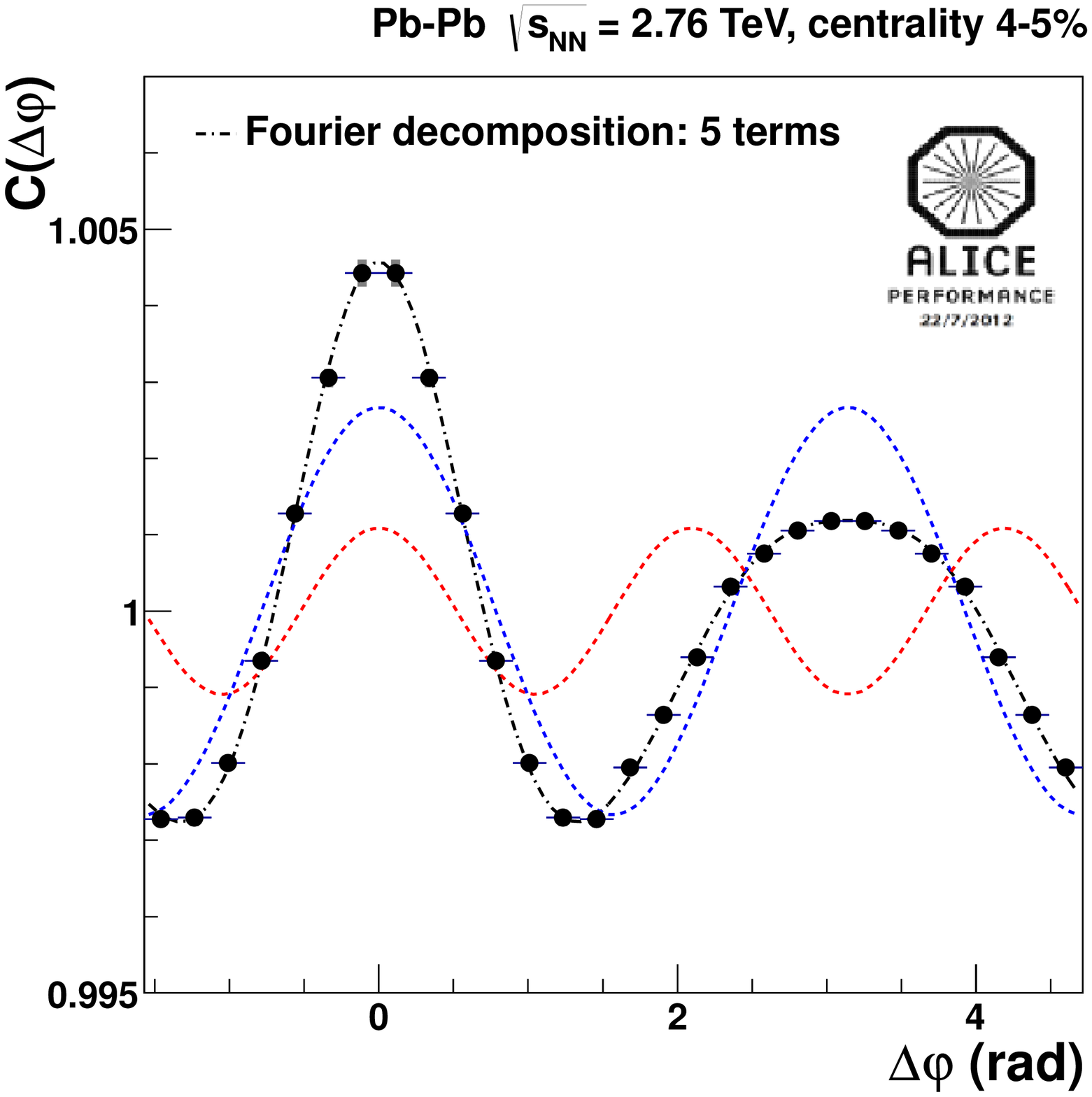}
&
\includegraphics[width = 0.46\textwidth]{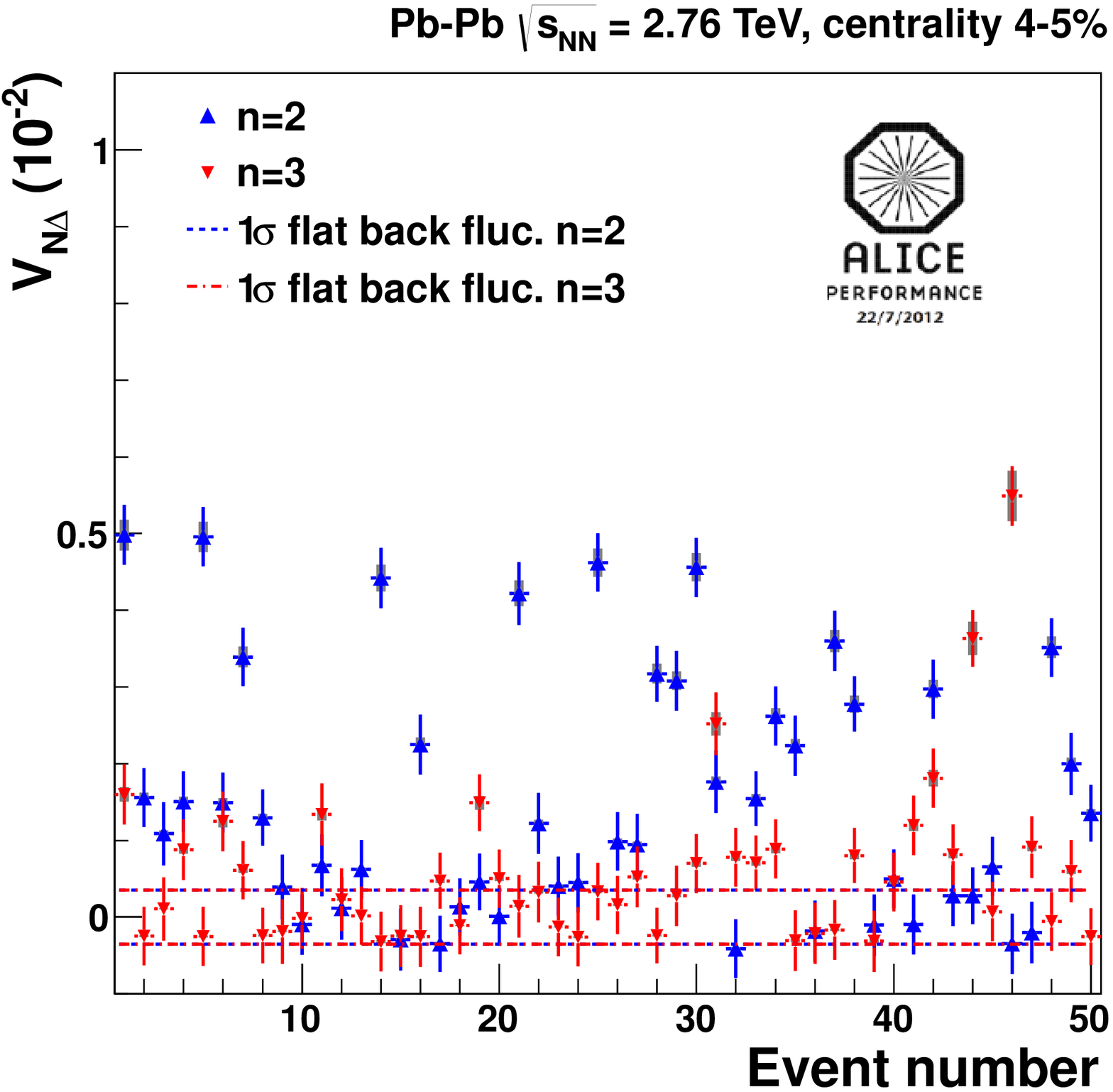}
\end{tabular}
\caption{Left Panel: Event averaged $\Delta \phi$ distributions for \Pb collisions. The dot-dash lines show the sum of a Fourier decomposition (with $n=2$ and $n=3$ contributions below). Right Panel: Event wise $V_{n\Delta}$ extracted from 50 events in chronological order. Dashed lines show $\pm 1\sigma$ statistical fluctuations for events with true $v_n=0$.}
\label{Figure1}
\end{figure}

\section{The correlation function}

We use charged hadrons with transverse momentum $p_{T} > 0.15$ GeV/c from the ALICE TPC to form the pair correlation function. This is defined as follows:
\begin{equation} 
\label{eq:C}
C(\Delta \phi)_{i} = \frac{N_{pairs} (\Delta \phi)_{i}} {\langle N_{pairs} \rangle}
\end{equation}  
where $N_{pairs} (\Delta \phi)_{i}$ is the number of pairs in bin $i$, and ${\langle N_{pairs} \rangle}$ is the average number of pairs summed over all bins.  It therefore quotes the relative deviation from the mean number of pairs. The main advantage of the correlation function is that pair inefficiencies due to the detector (e.g.  tracking inefficiencies) in the numerator and denominator cancel to leading order. To extract $V_{n\Delta}$ from $C(\Delta \phi)$, a definition used in a previous ALICE publication will be adopted \cite{HadDe}. It is given by:
\begin{equation} 
\label{eq:vNEx}
V_{n\Delta} = \sum_{i} C_{i} cos (n\Delta \phi) /  \sum_{i} C_{i}
\end{equation}  
where $n$ is the harmonic of interest and $C_{i}$ is the value of the correlation function in a particular $\Delta \phi$ bin. The key advantage of this method is that the calculation of $V_{n\Delta}$ is invariant with respect to the choice of normalisation of the correlation function. The left panel of figure \ref{Figure1} shows the event averaged correlation function obtained for Pb-Pb centrality 4-5\%. A Fourier decomposition is performed, and each coefficient is extracted using equation \ref{eq:vNEx}. The Fourier series is shown by the black lines. It was found that the first 5 terms are needed for a good description of the data for this $p_T$ range, and one can obverse the $n=2$ contribution is larger than the $n=3$ contribution.

\section{Event by event $V_{N\Delta}$}

Rather than obtaining $C(\Delta \phi)$ and $V_{n\Delta}$ from an event ensemble, we investigate these measurements in single events. In the right panel of figure \ref{Figure1}, event-wise values of $V_{2\Delta}$ and $V_{3\Delta}$ are shown for 50 events that occur for centrality of 4-5\% chronologically. The dashed lines show expected $\pm1\sigma$ levels of background fluctuations for each order, due to statistical fluctuations. There are clearly many events with $V_{2\Delta}$ higher than expected from just background fluctuations.  Such an observation indicates large fluctuations in the initial conditions, even within a small centrality range. The $V_{2\Delta}$ fluctuations are bigger than the $V_{3\Delta}$ fluctuations, and this is consistent with the fact $\langle V_{2\Delta} \rangle$ is bigger than $\langle V_{3\Delta} \rangle$ (shown in the left panel). This is also inconsistent with background fluctuations, where we find all orders of $n$ fluctuate equally (the red and blue dashed lines overlap). 
\begin{figure}[t]
\begin{center}
\includegraphics[width = 0.9\textwidth]{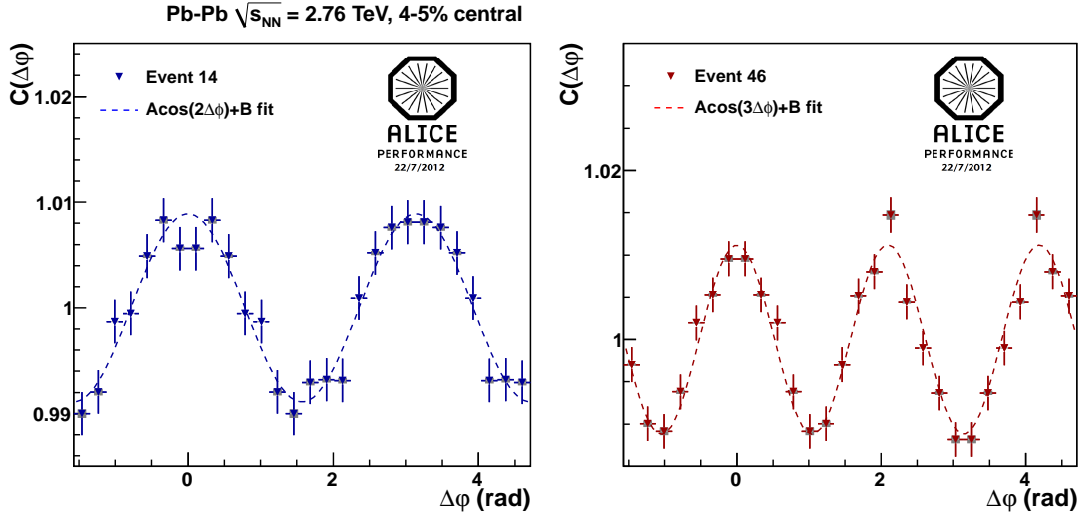}
\end{center}
\caption{$\Delta \phi$ distributions from events 14 and 46 in the right panel of figure \ref{Figure1}.}
\label{Figure2}
\end{figure}
In figure \ref{Figure2}, the $\Delta \phi$ distributions for event 14 and event 46 from the left panel of figure \ref{Figure1} are shown. It is observed that these events are completed dominated by elliptical ($n=2$) and triangular ($n=3$) flow respectively, and are in contrast with the distribution seen in the left panel of figure \ref{Figure1}.

\begin{figure}[h]
\begin{center}
\includegraphics[width = 0.97\textwidth]{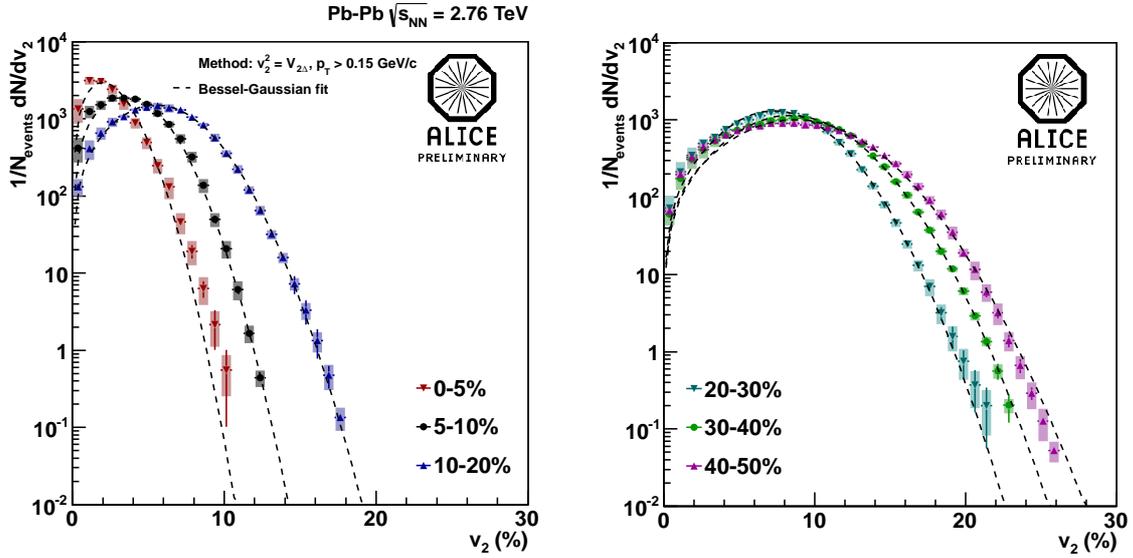}
\end{center}
\caption{Unfolded $v_{2}$ distributions as a function of centrality, subject to a $\Delta \eta > 0.8$ gap.}
\label{Figure3}
\end{figure}

\section{Unfolded $v_{2}$ distributions}

In figure \ref{Figure3}, we show $v_{2}$ distributions as a function of centrality. These are obtaining by unfolding raw $V_{2\Delta}$ distributions and applying the transformation $f(v_{2})=f(\sqrt{V_{2\Delta}})$, where $f(V_{2\Delta})$ is the unfolded $V_{2\Delta}$ distribution. The unfolding procedure \cite{unfold} is needed to remove the effects of the non-flow and trivial statistical smearing. The statistical smearing is determined by sampling a $dN/d\phi$ distribution with a known input $V_{2\Delta}$, then extracting the event-wise reconstructed $V_{2\Delta}$. The sampling is repeated, so one obtains the distribution of reconstructed $V_{2\Delta}$ for the given input. The contribution of non-flow is determined by fitting the pseudo-rapidly dependence of $V_{2\Delta}$ with a Gaussian (non-flow) and constant (flow). The systematic uncertainties (shown by the boxes) are due to the uncertainty in the non-flow contribution (evaluated by changing the $\Delta\eta$ cut), statistical smearing hypothesis (evaluated by changing the number of particles used) and centrality selection. The $v_{2}$ distributions are expected to reflect the participant eccentricity distributions \cite{ecc}, therefore provide constraints on models of the initial conditions. We fit these distributions with a Bessel-Gaussian distribution, which is the relevant p.d.f if the components of the eccentricity vector ($\varepsilon_{x}$,$\varepsilon_{y}$) are subject to Gaussian fluctuations \cite{GausFluc}. These fits generally describe the data well, bar central collisions. We found (not shown) that the unfolded $\langle v_{2}  \rangle$ and $\sqrt{\langle v_{2}^{2}  \rangle+2\sigma^{2}}$ are consistent with ALICE measurements of $v_{2}\{4\}$, and $v_{2}\{2\}$ respectively\cite{HigherOrder}, which is expected if the $v_{2}$ distributions are Bessel-Gaussian. The widths of the unfolded distributions were also found to be consistent with the widths obtained via $\sqrt{0.5(v_{2}\{2\}^{2}- v_{2}\{4\}^{2})}$.

\section{Summary}

We have investigated event by event di-hadron correlations, and found some events dominated by $n=2$ (elliptical events), and others dominated by $n=3$ (triangular events). The observations are beyond what is expected from trivial statistical fluctuations. We have presented unfolded $v_{2}$ distributions and found they are generally described by a Bessel-Gaussian distribution. Theses distributions can be used as constraints for models of initial state eccentricities.

\end{document}